\documentclass{emulateapj}
\usepackage{apjfonts}
\lefthead{HAN} 
\righthead{PLANETS WITH TWO SUNS}

\begin{document}
\title{Microlensing Search for Planets with Two Simultaneously Rising Suns}

\author{Cheongho Han}
\affil{Program of Brain Korea 21, Department of Physics,
Chungbuk National University, Chongju 361-763, Korea;
cheongho@astroph.chungbuk.ac.kr}



\begin{abstract}
Among more than 200 extrasolar planet candidates discovered to 
date, there is no known planet orbiting around normal binary 
stars.  In this paper, we demonstrate that microlensing is a
technique that can detect such planets.  Microlensing discoveries 
of these planets are possible because the planet and host binary 
stars produce perturbations at a common region around center of 
mass of the binary stars and thus the signatures of both planet 
and binary can be detected in the light curves of high-magnification 
microlensing events. The ranges of the planetary and binary 
separations of systems for optimal detection vary depending on 
the planet mass.  For a Jupiter-mass planet, we find that high 
detection efficiency is expected for planets located in the range 
of $\sim$ 1 AU -- 5 AU from the binary stars which are separated 
by $\sim$ 0.15 AU -- 0.5 AU.  
\end{abstract}

\keywords{gravitational lensing -- planets and satellites: general}


\section{Introduction}

Since the first discovery by \citet{wolszczan92}, more than 200 
extrasolar planet candidates have been discovered. Among the 
currently known extrasolar planet-hosting stars, approximately 20\% 
are members of binaries or multiple systems \citep{haghighipour06}. 
Nearly all of the planets in these systems revolve around only one 
of the stars. The only exception is the planetary system PSR B1620-26 
\citep{lyne88, backer93}, but this system consists of not normal 
stars but a pulsar and a white dwarf.

Besides the orbital configuration of the known planets in binaries 
where the planet revolves around one of the widely separated binary 
members, a planet can stay in a stable orbit if the planet is 
located well outside the binary stars and thus seeing the two stars 
as approximately a single object. However, such planets, which were 
imagined as a planet with two simultaneously rising suns in the 
{\it Star Wars} saga (Tatooine), have not been seen in planet searches 
using the radial velocity technique because the search programs avoid 
short-period binary stars. The chance to detect such planets via the 
transit technique is also very low because these planets tend to have 
wide orbits.

Planets in binary stellar systems can be detected by using the 
microlensing technique. Such a possibility was first mentioned by 
\citet{bennett99}.  They reported a candidate planet orbiting around 
a binary system from the observation and light-curve modeling of a 
microlensing event MACHO-97-BLG-41. However, their interpretation of
this event is very special in the sense that the source trajectory 
is precisely aligned with the line connecting the planet and 
the center of mass of the binary. Besides, the interpretation of 
the signal was rejected because a rotating binary-lens model 
provides an excellent fit to the observed data \citep{albrow00}. 
\citet{lee08} also investigated the lensing signal of planets in 
binary systems. However, their analysis is restricted to planets 
orbiting around one of the binary members. In this paper, we 
investigate the feasibility of detecting planets with two 
simultaneously rising suns by using the microlensing technique.


\begin{figure}[ht]
\epsscale{1.2}
\plotone{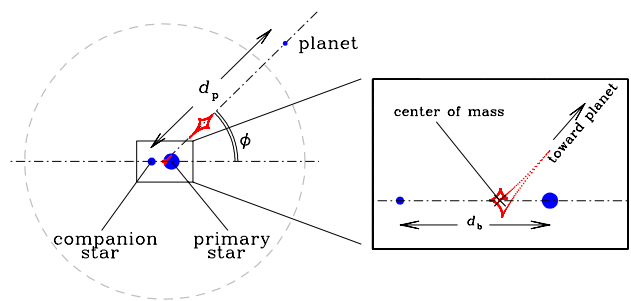}
\caption{\label{fig:one}
Lensing geometry of a planet revolving around close binary stars. 
The big dashed circle  represents the Einstein ring of a mass 
corresponding to the total mass of the binary stars. The angle 
$\phi$ denotes the position angle of the planet as measured 
counterclockwise from the binary axis and $d_b$ and $d_p$ denote 
the projected separations between the binary stars and between 
the planet and the binary center of mass, respectively. The right 
panel shows the enlarged view of the region around the binary stars. 
The figure drawn in red color represents the lensing caustic produced 
by the lens system.
}\end{figure}

\begin{figure*}[ht]
\epsscale{0.7}
\caption{\label{fig:two}
Lensing magnification patterns of triple lens systems composed of 
a planet revolving around binary stars. The labels above and on 
the right represent the projected separations between binary stars 
($d_b$) and between the planet and the center of mass of the binary 
($d_p$), respectively. In each map, the coordinates are centered 
at the position of the center of mass of the binary stars and the 
$x$-axis is aligned with the binary axis. The heavier star is on 
the right.  The planets have a common position angle of $\phi=
45^\circ$ as measured counterclockwise from the binary axis (see 
Figure~\ref{fig:one}).  The grey-scale is drawn such that brighter 
tone represents the region of higher magnification. The figures 
drawn in solid curves represent the lensing caustics. The presented 
patterns are for a Saturn-mass planet revolving around binary stars 
with a total mass of $0.5\ M_\odot$ and a companion/primary mass 
ratio of 0.5. The light curves resulting from the source trajectories 
marked by straight lines with arrows in the individual panels are 
presented in the corresponding panels of Figure~\ref{fig:three} 
(thick solid curves in black color).  The panels blocked by thick 
black solid lines show the cases where the planet dominates the 
perturbation pattern, while the panels blocked by thick white 
solid lines show the opposite cases where the binary stars dominate 
the pattern.
}\end{figure*}

\section{Lensing Behavior}

The general geometry of a planet revolving around the stars of 
a close binary is such that the separation between the stars is 
much smaller than the star-planet separation (see Figure 1). 
Under this geometry, the lensing behavior of the triple lens 
system can be greatly simplified because the close stellar binary 
pair and the planet can be separately treated.

The lensing behavior of the close stellar binary pair is mostly 
described by the single mass lens located at the center of the 
mass of the binary system with a mass equal to the total mass of 
the binary.  A small deviation occurs around the caustic which 
forms close to the center of mass.  The caustic represents the 
set of source positions at which the magnification of a point 
source becomes infinite.  For a close binary, the caustic has 
the shape of an asymmetric hypocycloid with four cusps and the 
asymmetry increases as the mass ratio between the stars, $q_b
=m_2/m_1$, decreases. Here $m_1$ and $m_2$ represent the masses 
of the heavier and lighter binary components, respectively.  
The caustic size depends weakly on the mass ratio, but it decreases 
rapidly with the decrease of the stellar separation.  For a close 
binary, then, the perturbation region is confined to a small 
region around the center of mass of the binary.

The planet, on the other hand, causes formation of two sets of 
disconnected caustics. One set is located away from the host star. 
The other set is located close to the planet-hosting star, which 
coincides with the region of the binary-induced perturbation. 
The latter caustic (central caustic) has an elongated wedge-like 
shape. The size of the planet-induced central caustics depends 
both on the star-planet separation and the planet/star mass ratio 
\citep{chung05}. The planet-induced caustic is small due to the 
small mass ratio of the planet. However, its size becomes 
non-negligible when the planetary separation is equivalent to 
the Einstein radius \citep{mao91, gould92}.

The small sizes of the deviation regions induced by the planet and 
the host binary stars greatly simplify the description of the 
lensing behavior of the system. This is because each deviation can 
be treated as a perturbation and thus the lensing behavior of the 
triple lens system is approximated as the superposition of those 
of the two sets of binary lenses composed of the binary star pair 
and the pair composed of the planet and a virtual star located at 
the center of the mass of the binary stars with a mass equal to 
the total mass of the binary. In addition, the coincidence of the 
individual perturbation regions might enable to detect both signatures
 of the planet and host binary stars from high magnification events 
for which the source trajectory passes close to the common region of 
perturbation.

\begin{figure*}[ht]
\epsscale{0.7}
\plotone{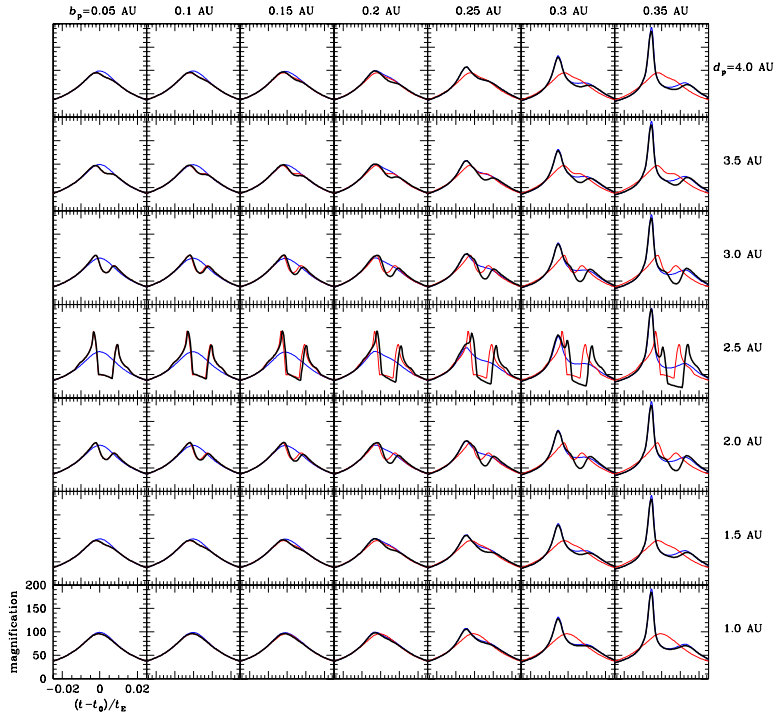}
\caption{\label{fig:three}
Light curves of lensing events caused by triple lens systems 
composed of a planet revolving around binary stars. The lens 
system geometry and the source star trajectories responsible 
for the individual events are presented in the corresponding 
panels of Figure~\ref{fig:two}. In each panel, there exist 
three light curves. The thick black solid light curve is for the 
triple-lens event, while the curves drawn in red and blue colors 
are for the cases where there is no planet and the planet-hosting 
star is a single object, respectively.
}\end{figure*}

\section{Magnification Pattern}

With the possible approximation of binary superposition in mind, 
we investigate the magnification pattern in the central region of 
various triple lens systems composed of a planet around binary 
stars. For this, we construct magnification map, which represents 
the map of lensing magnification as a function of the source star 
position.  For the construction of the map, we use the ray-shooting 
method.  
We choose this method because it allows not only simple procedure 
of multiple-lensing magnification computation but also easy 
incorporation of the finite-source effect.

Figure 2 shows the maps for a Saturn-mass planet revolving around 
binary stars with a total mass of $M=0.5\ M_\odot$ and a mass ratio 
of $q_b=0.5$.  The labels above and on the right side of the figure 
represent the projected separations between binary stars (binary 
separation, $d_b$) and between the planet and the center of mass of 
the binary (planetary separation, $d_p$), respectively. In each map, 
the coordinates are centered at the position of the center of mass 
of the binary stars and the $x$-axis is aligned with the binary axis, 
where the heavier star is on the right. For all tested cases, the 
planets have a common position angle  of $\phi=45^\circ$ as measured 
counterclockwise from the binary axis (see Figure~\ref{fig:one}). 
The grey-scale is drawn such that brighter tone represents the region 
of higher magnification. The figures drawn in solid curves represent 
the caustics.  Since the caustics induced by the planet and binary 
stars are small, finite-source effect would be important in 
magnification pattern \citep{bennett96}. We consider the effect by 
assuming that the source star has a radius equivalent to that of the 
Sun. We assume that events are observed toward the Galactic bulge 
field and the distances to the lens and source are $D_L=6$ kpc and 
$D_S=8$ kpc, respectively, by adopting those of a typical Galactic 
bulge event \citep{han95}. Then, the corresponding Einstein radius 
is $r_{\rm E}=(4GM/c^2)^{1/2}[D_L(D_S-D_L)/D_S]^{1/2}\sim 2.5$ AU 
and the source size normalized by the Einstein radius is $\rho=
1.5\times 10^{-3}$.

In Figure~\ref{fig:three}, we also present the light curves 
resulting from the source trajectories marked by straight lines on 
the maps. In each panel, there exist three light curves. The thick 
solid light curve is for the triple-lens event, while the curves 
drawn in red and blue colors are for the cases where there is no 
planet and the planet-hosting star is a single object, respectively.

From Figure 2 and 3, we find that the influence of the planet and 
binary stars on the pattern of perturbation varies depending on the 
binary and planetary separations. When the binary separation is small 
and the planetary separation is equivalent to the Einstein radius, 
the planet dominates the perturbation pattern. These correspond to 
the cases blocked by thick black solid lines in Figure 2. On the 
contrary, the binary dominates the perturbation pattern when the 
binary separation is considerable and the planetary separation is 
either too large or too small compared to the Einstein radius. 
These correspond to the cases enclosed by thick white solid lines 
in Figure~\ref{fig:two}. For events produced by these lens systems, 
then, the resulting perturbation would be well described by a binary 
lens of either star-planet or binary star pair. However, for lens 
systems with separations located in the neutral region of the 
$d_b$--$d_p$ parameter space, the influences of the planet and stellar 
binary are equivalent and thus it would be possible to detect both 
signals of the planet and binary stars.

\begin{figure}[ht]
\epsscale{1.2}
\caption{\label{fig:four}
The optimal ranges of binary and planetary separations for which 
microlensing technique is sensitive to the detections of planets 
orbiting around binary stars.
}\end{figure}

\begin{deluxetable}{ccc}
\tablecaption{Optimal Separation Ranges\label{table:one}}
\tablewidth{0pt}
\tablehead{
\multicolumn{1}{c}{planet} &
\multicolumn{1}{c}{planetary} &
\multicolumn{1}{c}{binary} \\
\multicolumn{1}{c}{type} &
\multicolumn{1}{c}{separation} &
\multicolumn{1}{c}{separation} 
}
\startdata
Jupiter-mass  &  $1.0\ {\rm AU}\lesssim d_{\rm p}\lesssim 5.0\ {\rm AU}$  & $0.15\ {\rm AU}\lesssim d_{\rm b} \lesssim 0.5\ {\rm AU}$ \\
Saturn-mass   &  $1.6\ {\rm AU}\lesssim d_{\rm p}\lesssim 3.8\ {\rm AU}$  & $0.15\ {\rm AU}\lesssim d_{\rm b} \lesssim 0.5\ {\rm AU}$ \\
\smallskip
Uranus-mass   &  $2.0\ {\rm AU}\lesssim d_{\rm p}\lesssim 3.0\ {\rm AU}$  & $0.15\ {\rm AU}\lesssim d_{\rm b} \lesssim 0.5\ {\rm AU}$
\enddata 
\end{deluxetable}

\section{Optimal Binary and Planetary Separations}

Then, what are the optimal ranges of the binary and planetary 
separations for which the microlensing technique is sensitive 
to the detections of planets orbiting around binary stars. 
In this section, we investigate these optimal ranges.

The size of the perturbation region is proportional to the 
size of the caustic. We, therefore, determine the optimal 
region in the $d_b$--$d_p$ parameter space as the one within 
which the sizes of the planet and binary-induced caustics 
become equivalent.  Figure~\ref{fig:four} shows the regions 
found under this approximation for three different types of 
Jupiter, Saturn, and Uranus-mass planets.  For lens systems 
with planetary and binary separations located in the shaded 
region, the size ratio between the two types of caustic is 
in the range of $1/5\leq \Delta\xi_p/\Delta\xi_b\leq 5.0$, 
where $\Delta\xi_p$ and $\Delta\xi_b$ represent the sizes 
of the planet and binary-induced caustics, respectively. 
We apply two exceptions to this rule. The first is that if 
both caustics are too small, we assume that the signal 
detection would be difficult due to severe finite-source 
effect. To account for the finite-source effect, we set the 
lower limit of the caustic size for signal detection as two 
times of the source size. The second exception is that if 
both caustics are greater than a certain value, we assume 
that the chance to detect both planet and stellar binary is 
not small. We set this threshold caustics size as 10 times 
of the source size. From the comparison of the optimal ranges 
determined in this way with those of the neutral region judged 
based on eye inspection of the maps in Figure~\ref{fig:two}, 
we find good agreement. The optimal ranges vary depending on 
the planet mass.  For a Jupiter-mass planet, we find that 
high detection efficiency is expected for planets located in 
the range of $1\ {\rm AU} \lesssim d_p \lesssim 5\ {\rm AU}$ 
orbiting around binaries with separations of $0.15\ {\rm AU}
\lesssim d_b\lesssim 0.5\ {\rm AU}$.  The range of the planetary 
separation shrinks as the planet mass decreases. However, the 
range of the binary separation remains nearly the same regardless 
of the planet mass.  In Table~\ref{table:one}, we summarize the 
optimal ranges for various types of planets.

It might be that planets located in some portions of the tested
$d_b$--$d_p$ parameter space might be dynamically unstable and 
thus the estimated optimal region is overestimated. At present, 
however, the possible extent of the planet separation around 
stars of a close binary system is an open question and is 
subject to be tested from observation.  In this sense, the 
constraint of the stable region of the planetary separation 
that will be provided by the microlensing method is very 
important.

\section{Discussion}

Interpreting the signals of triple-lens systems is not an easy 
task. With more complete understanding of lensing behavior and 
the development of efficient modelling codes, however, such 
analysis is being started for the analysis of observed lensing 
events (A. Gould private communication). Not long ago, binary-lens 
modelling was considered to be a difficult task, but now it is 
routinely conducted for binary events. Considering this, 
triple-lens modelling would be possible in future lensing analysis.

Nevertheless, one might think of several possible degenerate 
cases to the signal of the planet orbiting a close binary stars. 
One possible case is a triple lens system where a planet orbits 
around one of widely separated binary stars.  In this case, 
the binary-produced caustic is located also close to the 
planet-induced caustic, causing potential degeneracy.  However, 
the caustic produced by the wide-separation binary has a very 
symmetric hypocycloid shape, while the shape of the caustic 
produced by close binary stars is in general asymmetric.  
Therefore, the resulting patterns of the two cases are different 
and thus it would not be difficult to distinguish the two cases.  
Another possible case is a star with multiple planets \citep{gaudi98}.  
Here the perturbations in the central region are produced by the 
individual planets.  The planet-induced caustics are much more 
elongated than binary-induced caustic. Therefore, it would not 
be difficult to distinguish this case, either.

Planets around close binary stars will be detected through the 
channel of high-magnification events. Current microlensing 
follow-up observations \citep{beaulieu06, gould06} are focusing 
on these events due to their high efficiency in planet detections 
\citep{griest98}. By using multiple telescopes located at different 
locations, these observations enable dense and continuous coverage 
of events which is required to detect the short-lasting signature 
of the planet in binary stars. In addition, greatly enhanced source 
brightness of high-magnification events enables precision photometry 
which is essential for the characterization of the signature. 
Therefore, we predict that planets with two simultaneously rising 
suns would be detected in the near future.

\section{Conclusion}

We investigated the feasibility of detecting planets orbiting 
close binary stars by using the microlensing technique.  We 
presented the channel for which one can identify the signatures 
of such planetary systems.  We illustrated the signatures for 
various planet-binary configurations and explained the tendencies 
of the signatures.  We also estimated the ranges of the binary 
and planetary separations for which the microlensing method is 
efficient in detecting these planetary systems.  Discovering 
planets orbiting around close binary stars is difficult by 
using the radial velocity or transit technique.  Therefore, 
microlensing is an important technique that can provide 
information about these planetary systems, which is very 
important for our understanding of planet formation.

\acknowledgments 

This work was supported by the Science Research Center (SRC) 
program.


\clearpage

\end{document}